# SIMO channel performance evaluation on indoor environment at 2.4 GHz

Constantinos I. Votis, Vasilis Christofilakis* and Panos Kostarakis

*Physics Department, Electronics-Telecommunications and Applications Laboratory, University of Ioannina Panepistimioupolis, Ioannina, Greece*

* Corresponding author. Email: vachrist@uoi.gr

# SIMO channel performance evaluation on indoor environment at 2.4 GHz

This work presents an experimental study of Single Input Multiple Output (SIMO) channel performance in indoor radio propagation environment. Indoor channel measurements at 2.4 GHz ISM frequency band have been performed using a versatile channel sounder testbed platform. A single transmitting antenna, four receiving antennas with two proposed geometries and a four-branch receiver circuitry were used in order to achieve channel sounder measurements exploiting baseband signal processing techniques. Deep investigation on SIMO wireless channel performance was realized through three types of metrics which are signal strength, gain coefficient and capacity. Performance results indicate SIMO channel capacity enhancement and illustrate differences between the two proposed geometries.

**Keywords:** SIMO processing; channel capacity; antenna array configuration; test bed platform

## 1. Introduction

Nowadays, multiple-element wireless technologies on transmitter and/or receiver end promise channel capacity enhancement and remarkable spectral efficiency. In an indoor environment, such technologies ensure quite low correlated propagation paths that exist on the wireless channel between transmitter and receiver (Chua, Tse, Kahn and Valenzuala 2002; Da-Shan, Foschini, Gans and Kahn 2000; Saunders and Aragón-Zavala 2007).

SIMO architecture uses multiple-elements only on the receiver end. A single transmitting antenna radiates an RF signal, and the corresponding signal replicas are collected by an antenna array on the receiver end. In this way, SIMO architecture has the potential to increase the channel capacity through receiving diversity techniques (Votis, Tatsis, Christofilakis and Kostarakis 2012). It is well known that significant channel capacity increment is obtained by using multiple antennas on both transmitting and receiving sides creating a so-called MIMO architecture (Winters 1987; Foschini & Gans 1998). On the other hand, MIMO channel sounder needs a multiplexing technique to separate signals from all transmitting antennas, increasing the complexity and the cost of a communication system (Kim, Jeon, Lee, and Chung 2007). In this work, a SIMO channel sounder platform was developed instead of MIMO due to cost and simplicity reasons.

Extensive research on multi-antenna channel performance evaluation includes theoretical studies, realistic channel simulation models and mathematical manipulations on channel conditions (Da-Shan, Foschini, Gans and Kahn 2000; Hui, Yong, & Toh 2010; Mangoud 2012; Wang and Hui 2011). Production of commercial multi-antenna systems in indoor/outdoor environments' requires full knowledge and understanding of channel conditions, which can only be achieved with in-depth analysis of channel measurement data (Maharaj, Wallace, Jensen, & Linde 2008). In any case, research on multi-antenna architecture performance includes numerous channel characterization works that utilize theoretical and experimental aspects on various radio propagation environments. Some of them are based on theoretical processes (Golden, Foschini, Valenzuela, & Wolniansky 1999), some other are stochastic (Weichselberger, Herdin, Ozcelik & Bonek 2006 ; Bonek & Weichselberger 2005) and the rest of them depend on experimental measurements (Swindlehurst, German, Wallace & Jensen 2001 ; Martin, Winters, & Sollenberger 2001). These works exploit methods and techniques in order to confirm the significant channel capacity enhancement that is promised by using MIMO architecture. Both theoretical and stochastic works on MIMO channel characterization have no dependence on the exact characteristics of the radio propagation environment. Instead, experimental channel sounder measurements provide a versatile and efficient way in studying and investigating MIMO channel performance. Channel sounder platforms are utilized in order to provide experimental channel characterization. The resultant data offer a large amount of considerations that can provide channel propagation model development. Many useful methods are also introduced to enhance channel sounder applications such as space-time coding and OFDM techniques (Chronopoulos, Tatsis, Raptis & Kostarakis 2011).

In this paper, we describe experimental results on SIMO channel capacity enhancement that are obtained with a 1x4 SIMO channel sounder platform at 2.4 GHz. The present work is primarily motivated by the following facts:

- Concerning multiple receiver elements measurements, mainly uniform linear array configurations consists of omnidirectional monopoles or dipoles have been evaluated (Keignart, Abou-Rjeily, Delaveaud, & Daniele 2006; Nishimori, Kudo, Honma, Takatori & Mizogughi 2010). In the literature, few existing multiple elements systems employ uniform circular arrays can be found (Wallace & Jensen 2005; Mangoud & Mahdi 2011). In general planar array configurations will be a preferable selection for future generation WLANs due to their enhanced azimuth coverage (Mangoud 2012). In this work, we study the performance of the typical uniform linear array configuration compared with a new planar geometry forming the letter «Π».
- Furthermore, prototype printed antenna elements of low cost; compactness and reliability were used. These elements have already been tested to meet certain performance and high efficiency at the system's operation frequency of 2.4 GHz (Votis, Kostarakis, & Alexandridis 2010).
- Instead of semi or fully switched systems a full parallel SIMO channel sounder was developed that significantly reduces the time required to perform a sampling of the channel; the sounder can be successfully used in more environments (Laurenson & Grant 2006). It also offers an adaptable receiving antenna system in which a wide variety of different antenna arrays configurations could be measured.

- The proposed SIMO set-up uses an RF signal generator and a digital oscilloscope in order to achieve channel sounder measurements. In that way, the utilization of the proposed SIMO platform exhibits an extra motivation on channel sounder applications using a cost effective instrumental equipment of a typical electronics and telecommunications laboratory.

The rest of the paper is organized as follows. Section 2 presents the channel sounder testbed platform. The capacity estimation formula is introduced and analyzed in the first paragraph of Section 3. At the second paragraph of Section 3, the channel sounder measurements parameters are depicted and discussed. In this section, experimental measurements are also presented and discussed. The paper concludes in Section 4.

## 2. Channel sounder platform

A testbed set-up was designed, implemented and installed to provide an efficient and flexible investigation on SIMO channel capacity and channel characterization. This measurement platform provides SIMO channel study on indoor radio propagation environment. The construction of the proposed sounder platform exhibits quite cost effective performance as it necessitates common laboratory instrumentation. Having identical transmitting and receiving antenna architecture along with the multiple channel receiving architecture that consists of identical electronics ICs and components the cost effectiveness is obviously enhanced. Also, it offers a versatile way for full parallel SIMO channel sounder applications, without introducing various performance degradations due to using of switching devices on the receiver side such as time delays e.t.c. Those benefits far outweigh any possible disadvantage that may be related to the lack of real-time channel data manipulation or to the quite unwieldy antenna structure. The last could probably be a vital benefit in channel sounder applications that necessitate various multiple elements antenna geometries. Concerning the transmitter side a prototype printed dipole antenna with integrated balun on a reflector structure was used. The prototype printed antenna element has already been tested to meet certain performance and high efficiency at the frequency range of 2.4 GHz (Votis, Christofilakis, & Kostarakis 2010). Additionally, this printed dipole antenna architecture was introduced for constraints of low cost, compactness and reliability (Edward & Rees 1987; Garg, R. 2001; Railton & Hilton 2002; Chuang & Kuo 2003; Balanis 2005).

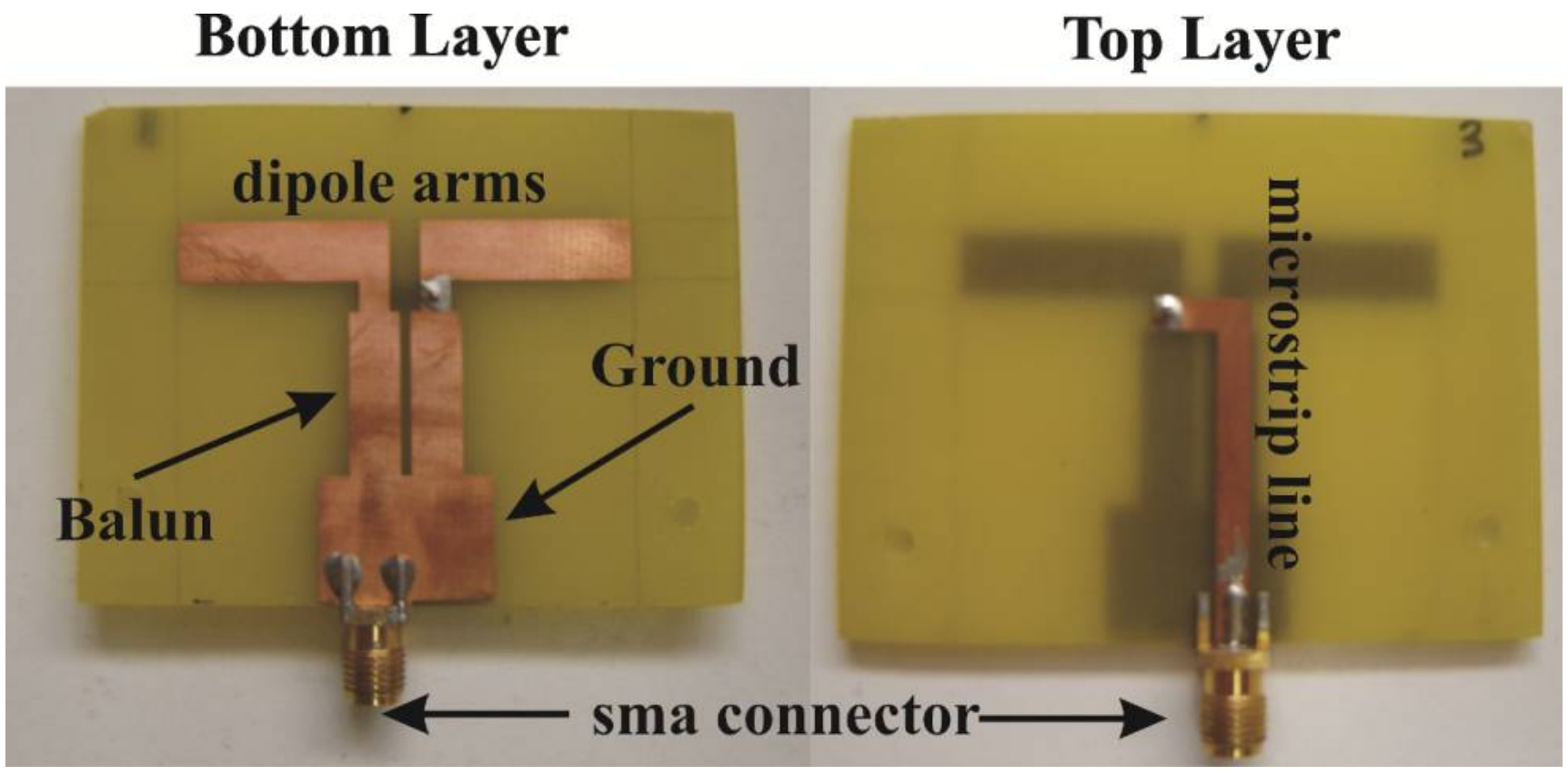

Figure 1. Printed dipole architecture

Bottom and the top layer of the printed dipole with integrated balun structure are shown in Figure 1. On the receiver side, the same printed dipole was used as an antenna element. Two receiving antenna array configurations consist of four (N=4) identical printed dipole antennas were proposed. The first one is a well known Uniform Linear Array (ULA) consists of four identical dipoles on a linear arrangement. The distance between two sequential elements approximates the half of the wavelength at the frequency range of 2.4 GHz (Figure 2). The proposed ULA performance was in detail investigated and discussed in terms of antenna element return loss and radiation pattern (Votis, Kostarakis, & Alexandridis 2010). The second one is the Π-shape antenna array configuration consists of four identical dipoles with an arrangement that resembles the shape of the Greek language character Π (Figure 3).

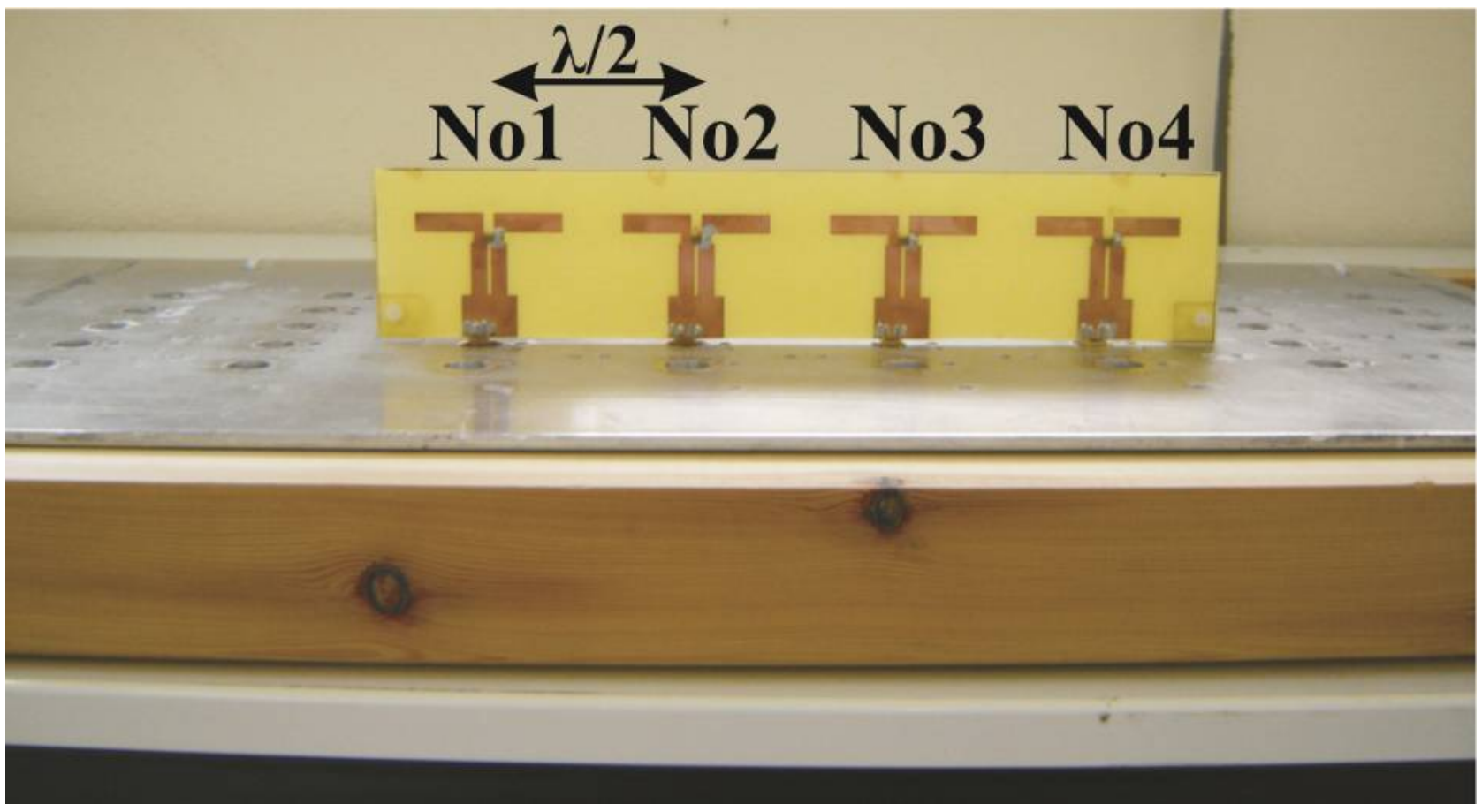

Figure 2. ULA configuration

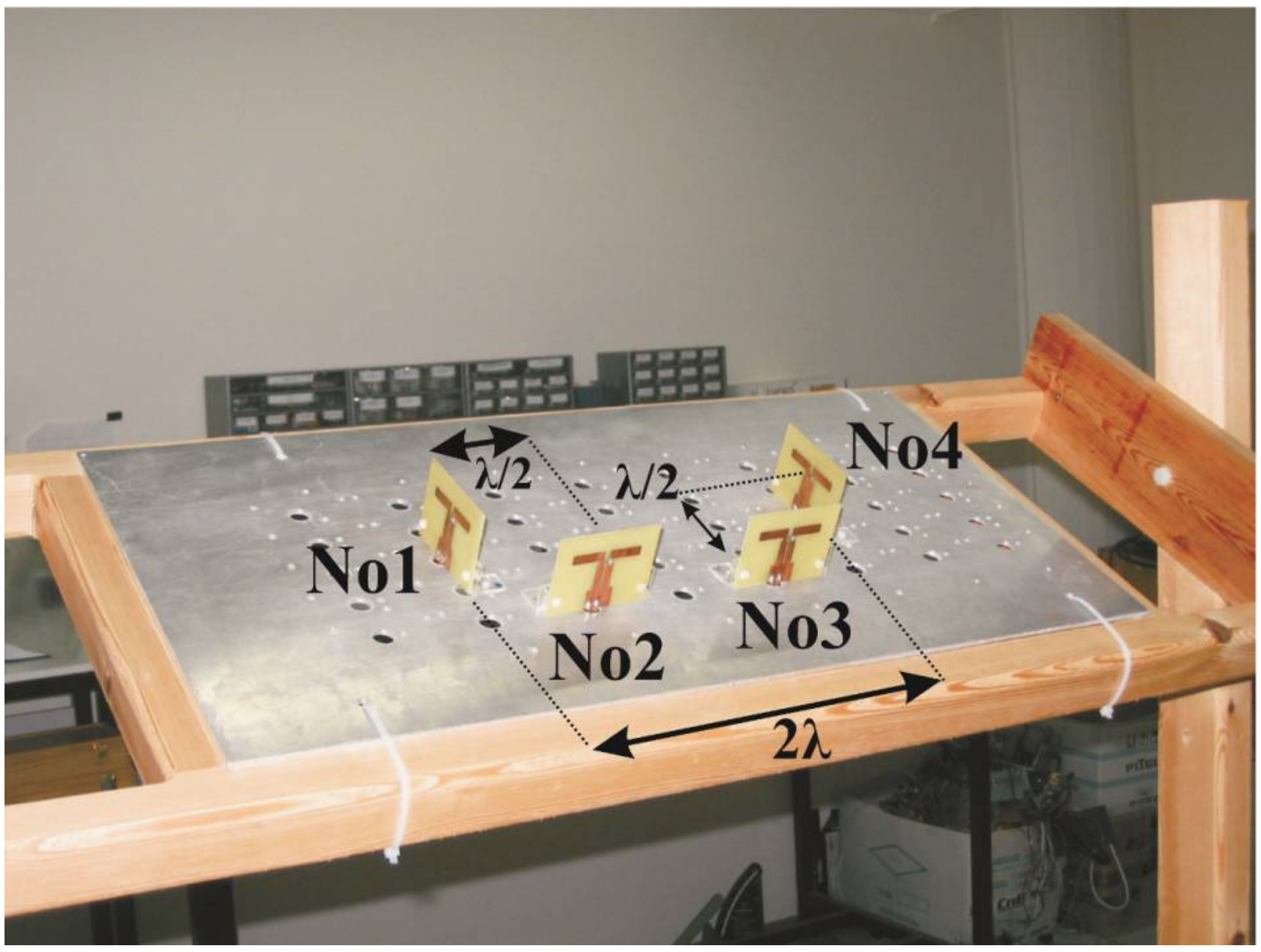

Figure 3. Π-shape array configuration

All the antenna elements and the reflector plates are placed on Specific Wooden Positioning Structures (SWPS) at transmitter and receiver communication link. SWPS, as shown in Figure 4, make available not only mechanical support but also provides antenna array rotation on both azimuth and elevation axis. The whole implementation was designed and constructed with no other metallic material used instead of the aluminium reflector plate, in order to eliminate antenna dipole performance degradation. In Figure 4, the adaptation of the ULA configuration and the reflector plate at the wooden positioning structure are shown.

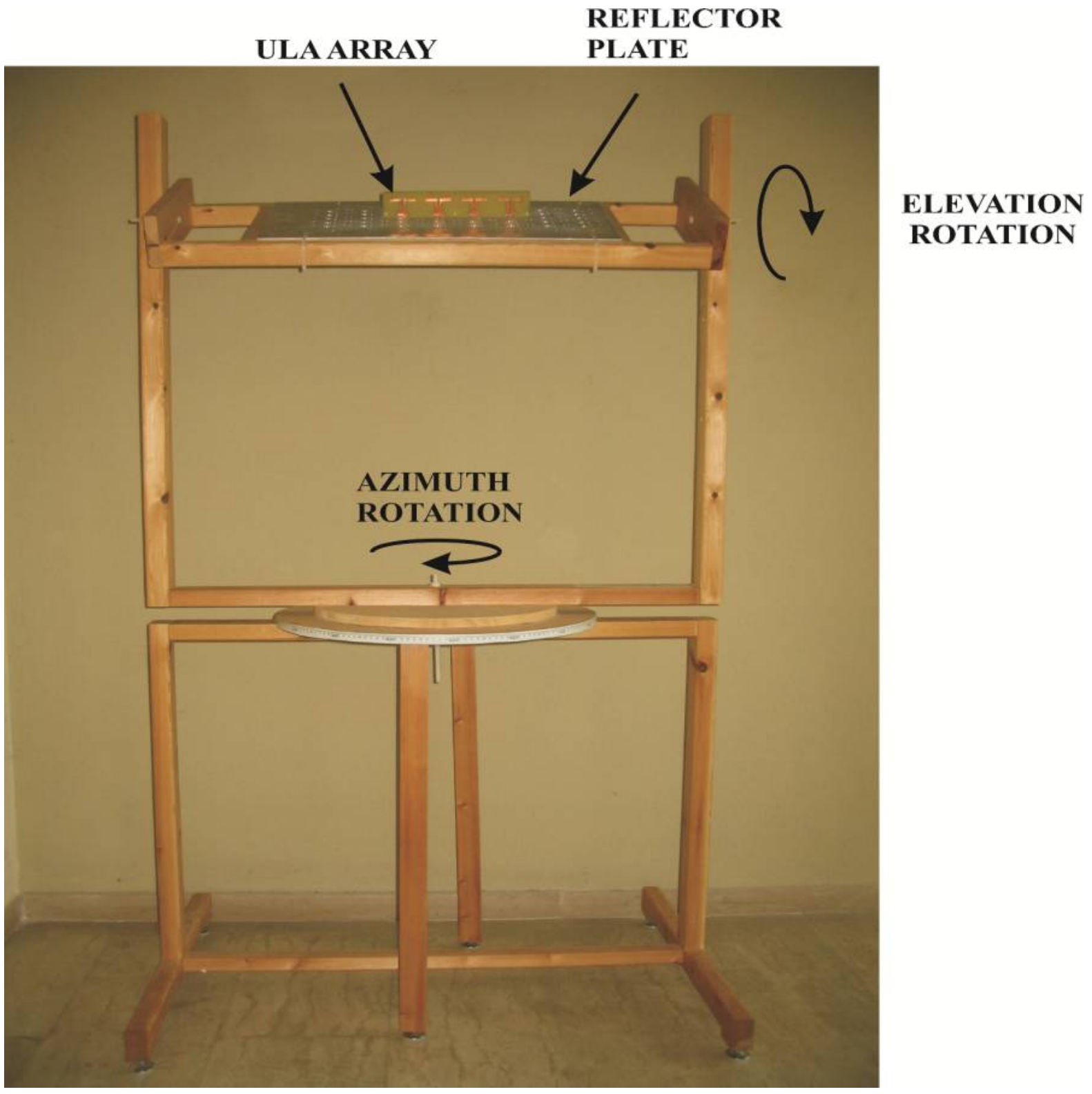

Figure 4. Specific Wooden Positioning Structure (SWPS)

The measurement platform was set-up in a typical rectangular laboratory environment with dimensions of 9m width by 12m length. Electronics-Telecommunications and Applications Laboratory (ETA Lab) plan is shown in Figure 5a. A schematic block of the test measurement platform is shown in Figure 5b. Except the 2 SWPSs (one for the Tx and one for the Rx end) placed at a distance 4.5m apart it contains: The transmitter signal generator that excites the one element at the transmitting side. At the receiver side, there is a four element antenna array configuration, the four branch receiver device, the receiver signal generator and a digital oscilloscope.

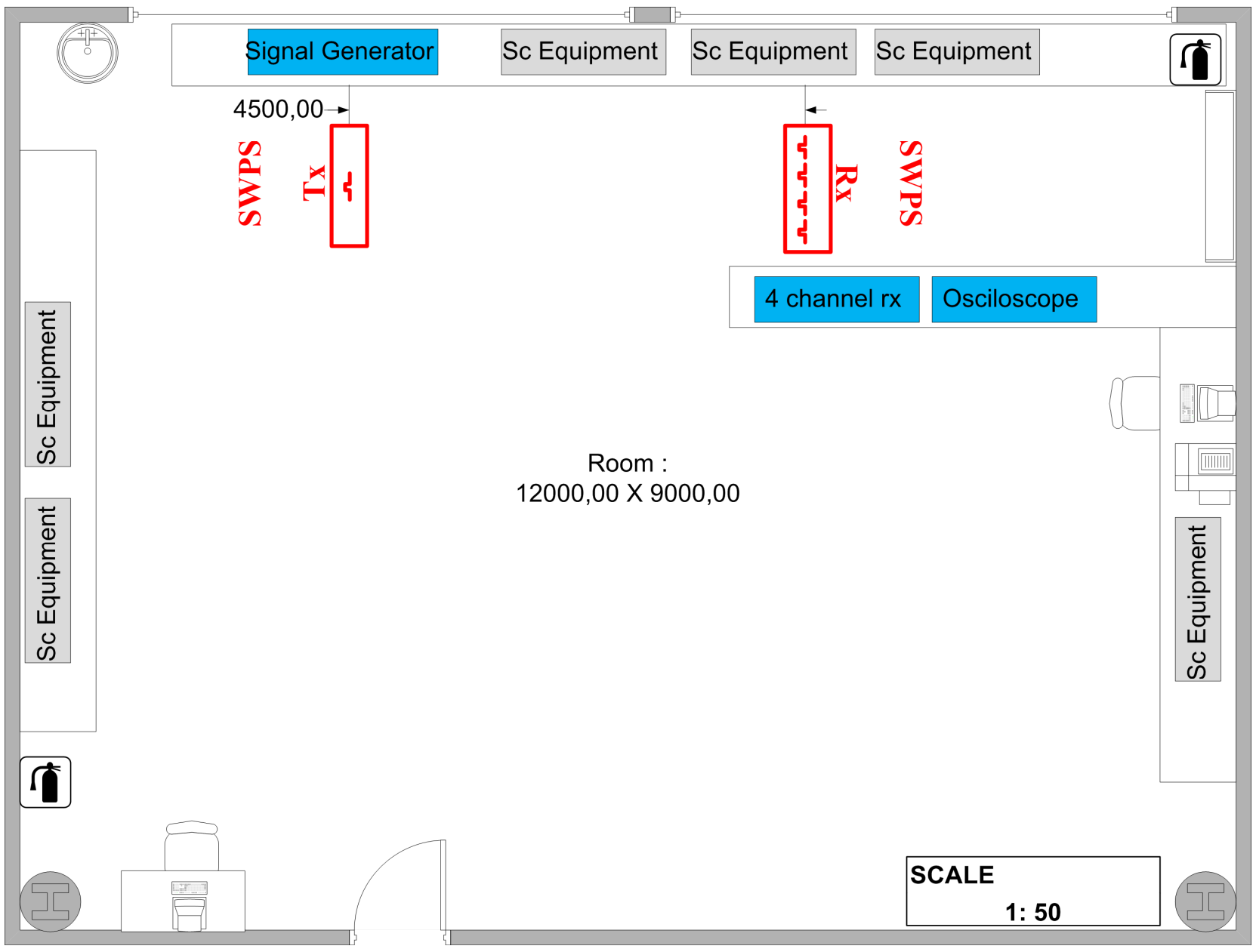

(a)

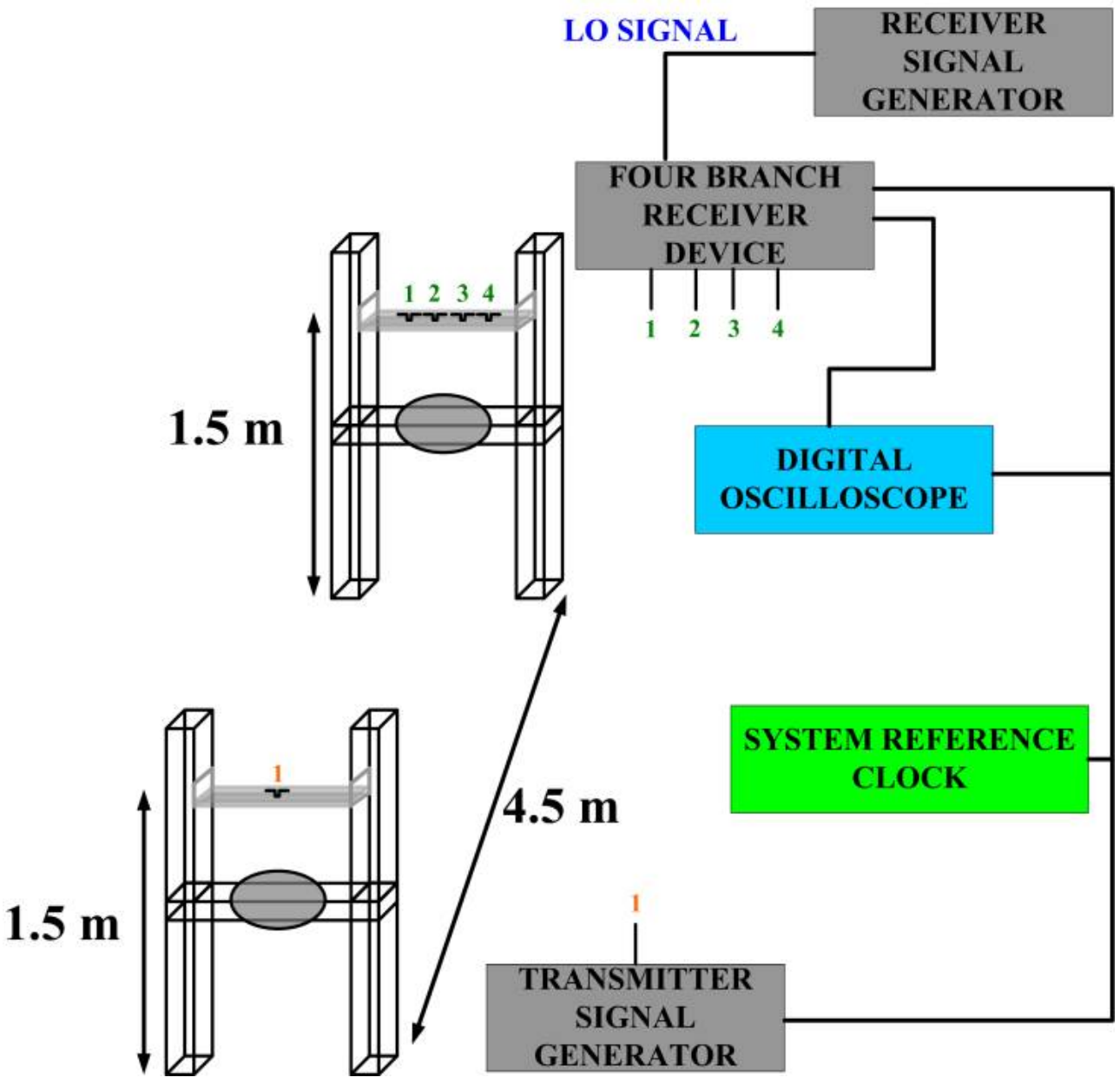

(b)

Figure 5. (a) ETA Lab Plan, (b) Test measurement platform,

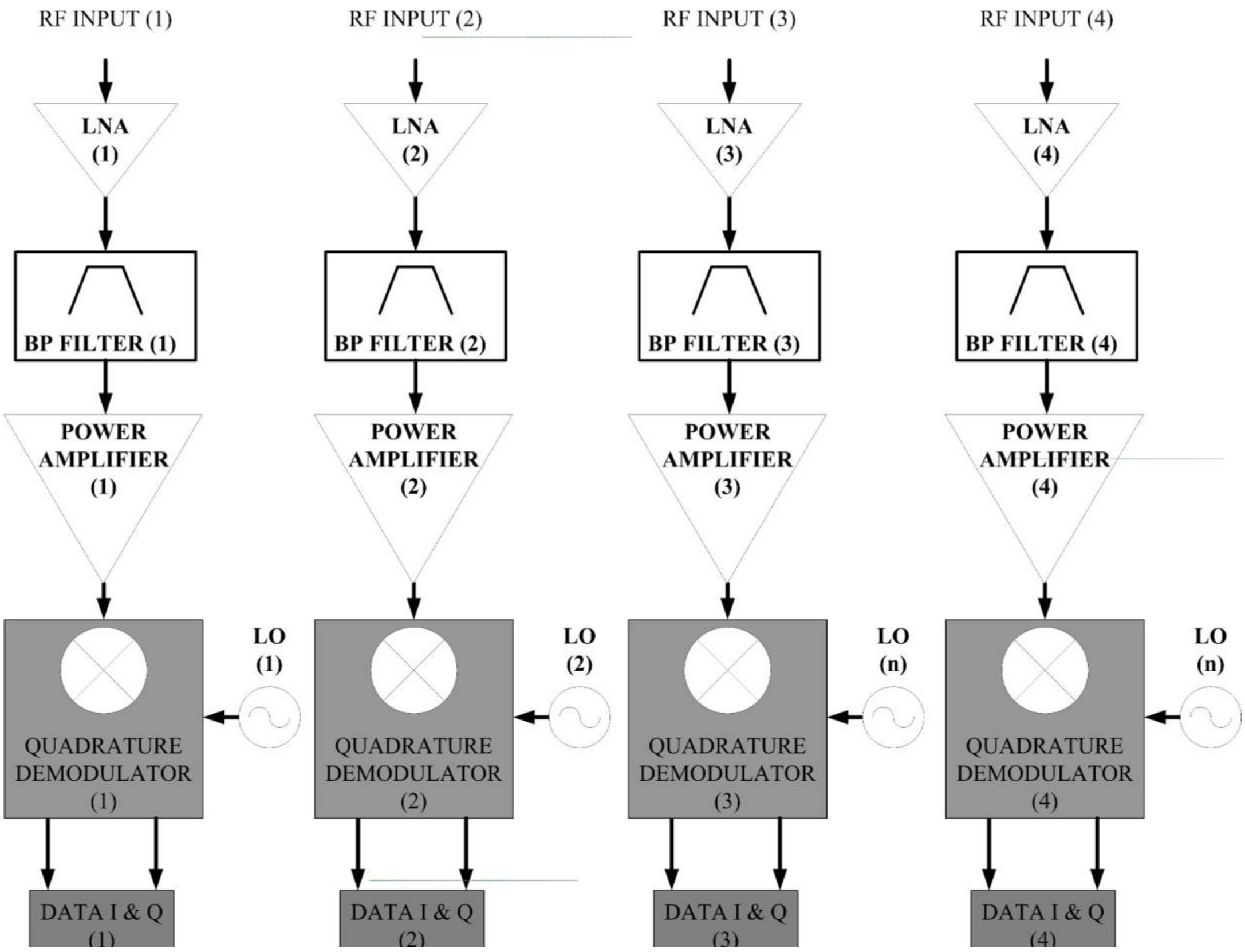

Figure 6. Four branch receiver architecture

Demodulation, amplification and filtering procedures were obtained by the four channel receiver. The receiver output signals are sampled and recorded by the digital oscilloscope that also exploits versatile data acquisition performance. From Figure 6, the receiving signal at each channel is firstly amplified by a low noise amplifier ("300MHz to 2500MHz," 2007) and then is filtered by a 2.45 GHz bandpass filter. Power signal strength enhancement on receiving signal is mainly achieved through power amplifier ("Monolithic Amplifier," 2011) at the output of the bandpass filter. That proposed receiver design exhibits a 42 dB single channel power gain at pass band frequency range. The AM to PM conversion approximates to 0.2 deg / dB at the mean value of receiving signal strength (- 62.5 dBm). The system reference clock provides the synchronization on channel sounder platform performance. The resultant data provide receiving signal strength, channel gain coefficients and channel capacity estimation using non-real-time baseband signal processing at a personal computer. Similar methods were proposed by Martin, Winters & Sollenberger 2001.

## 3. Measurements process analysis and results

### *3. 1 Channel sounder measurement parameters*

As already mentioned, all the measurements took place in a typical laboratory room where: The single transmitting dipole was excited by a single-tone radio signal at the frequency range of 2.4 GHz, using a laboratory RF signal generator and the power level of the transmitted RF signal was closed to - 8 dBm.

The experimental channel sounder measurements were referred to the same indoor laboratory propagation environment at two independent time intervals that had equal duration. Each experimental channel sounder time interval includes 100 sequential time snapshots. The time interval between two sequential measurement snapshots approximates to 4 ms due to inherent digital oscilloscope performance. In the case of static indoor radio propagation environments, the duration of 4 ms has no impact on channel characterization measurements (Rappaport 1996). It was also assumed AWGN and negligible frequency dependence of the wireless channel. This describes the narrow band assumption providing that the channel has frequency flat response. Wireless propagation environment was also characterized by the block fading model, providing that the channel complex gain coefficients have constant values for each data block time interval. The latest assumption ensures that the channel parameters for two sequential data blocks are quite uncorrelated. Channel sounder measurement parameters are summarized in Table 1.

Table 1. Channel sounder measurements parameters.

Channel Noise : Additive White Gaussian Noise

Channel response : Flat frequency

| Measurement parameter | Value |
|---|---|
| Time snapshots duration | 100 |
| Transmitter – Receiver distance | 4.5 m |
| Transmitter – Receiver antenna heights | 1.5 m |
| Time interval between sequential measurement snapshot | 4 msec |
| Transmitted single-tone power signal | - 8 dBm |

The 1x4 SIMO channel can be seen as four parallel sub-channels that co-existing in the wireless indoor environment. Each sub-channel is defined by the single transmitting antenna element and one of the four receiving antenna elements. In order to perform channel characterization we had to calculate the complex channel gain coefficient for each one of the four sub-channel. These four complex channel gain coefficients were calculated at each measurement snapshot every 4 ms. The resultant 1x4 matrix describes the temporal complex channel gain matrix defined as h. Each vector element $h_i$ is a complex number and, the i index ranges from 1 to 4. The channel capacity of SIMO wireless channel is then given by:

$$C = \log_2(\det(\mathrm{I} + \rho \cdot \mathrm{h}^{\dagger} \cdot \mathrm{h})) \qquad (1)$$

Where det() denote the determinant, I corresponds to the identity matrix and ρ is the signal to noise ratio at each receiver antenna element. The superscript denotes complex conjugate transpose (Martin, Winters & Sollenberger 2001).

Regarding the evaluation of the channel capacity enhancement due to the use of multiple antenna elements on the receiver end the following equation was used:

$$C_n = \frac{\log_2(\det(\mathrm{I} + \rho \cdot \mathrm{h}^{\dagger} \cdot \mathrm{h}))}{\frac{1}{4}\sum_{i=1}^{4} \log_2(1 + \rho \cdot (\mathrm{h}_{\mathrm{i}})^2)} \qquad (2)$$

The parameter $C_n$ denotes the normalized capacity. The denominator corresponds to the average capacity. Consequently, the ratio $C_n$ is the estimated capacity, averaging to all 4 SISO measured sub-channels.

### *3. 2 Channel sounder measurement results*

In the following subsections, deep investigation of the channel performance is presented and analyzed through different types of metrics

#### *3.2.1 Received signal strengths*

We firstly evaluated the 1x4 SIMO channel received signal strengths at two distinct time intervals (first and second). Figures 7 and 8 plot the received signal strengths for the ULA configuration at the first and the second time interval, respectively. Figures 9 and 10 plot the received signal strengths for the Π-shape array configuration at the first and the second time interval, respectively.

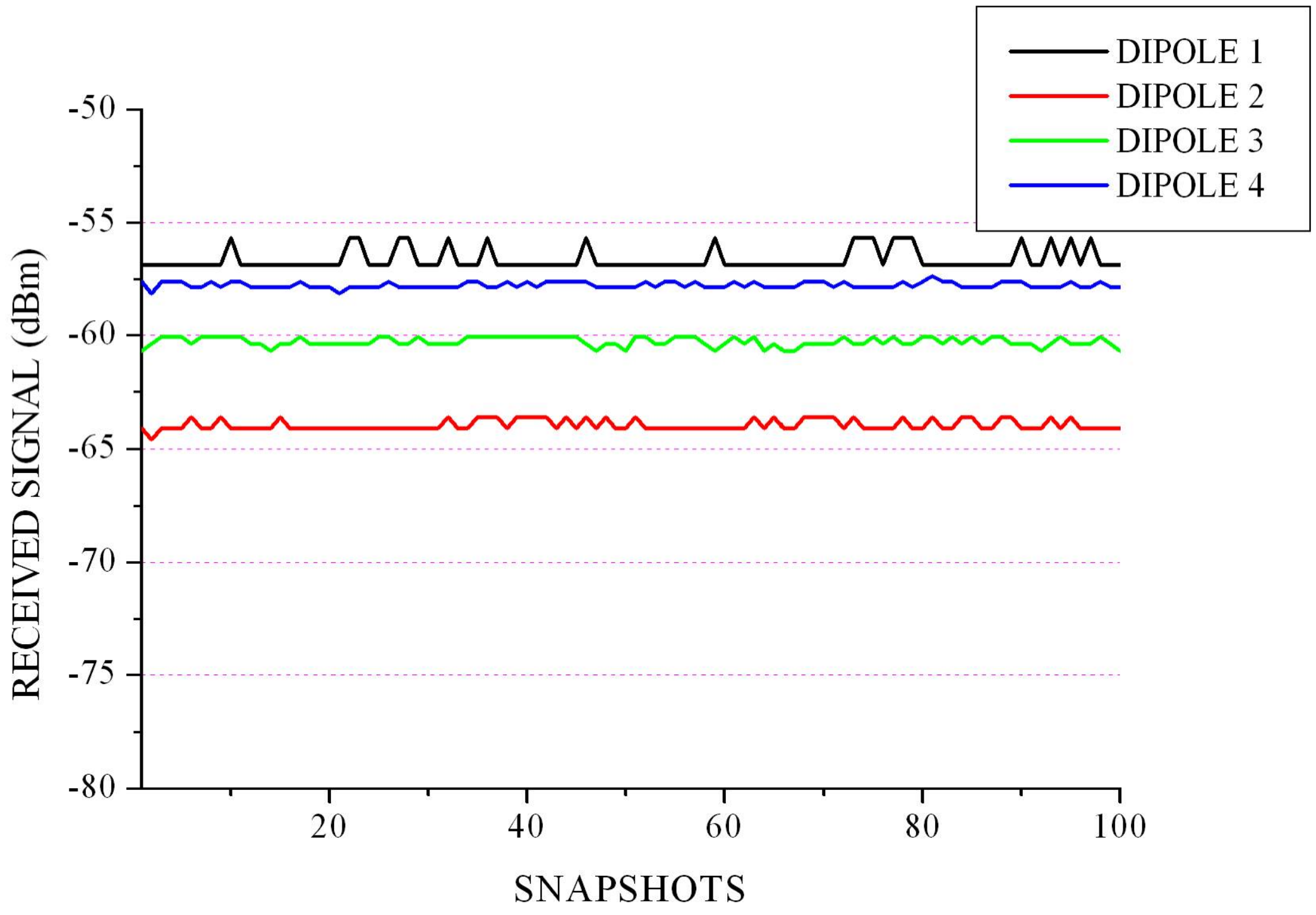

Figure 7. Received signal strength on ULA at the first time interval

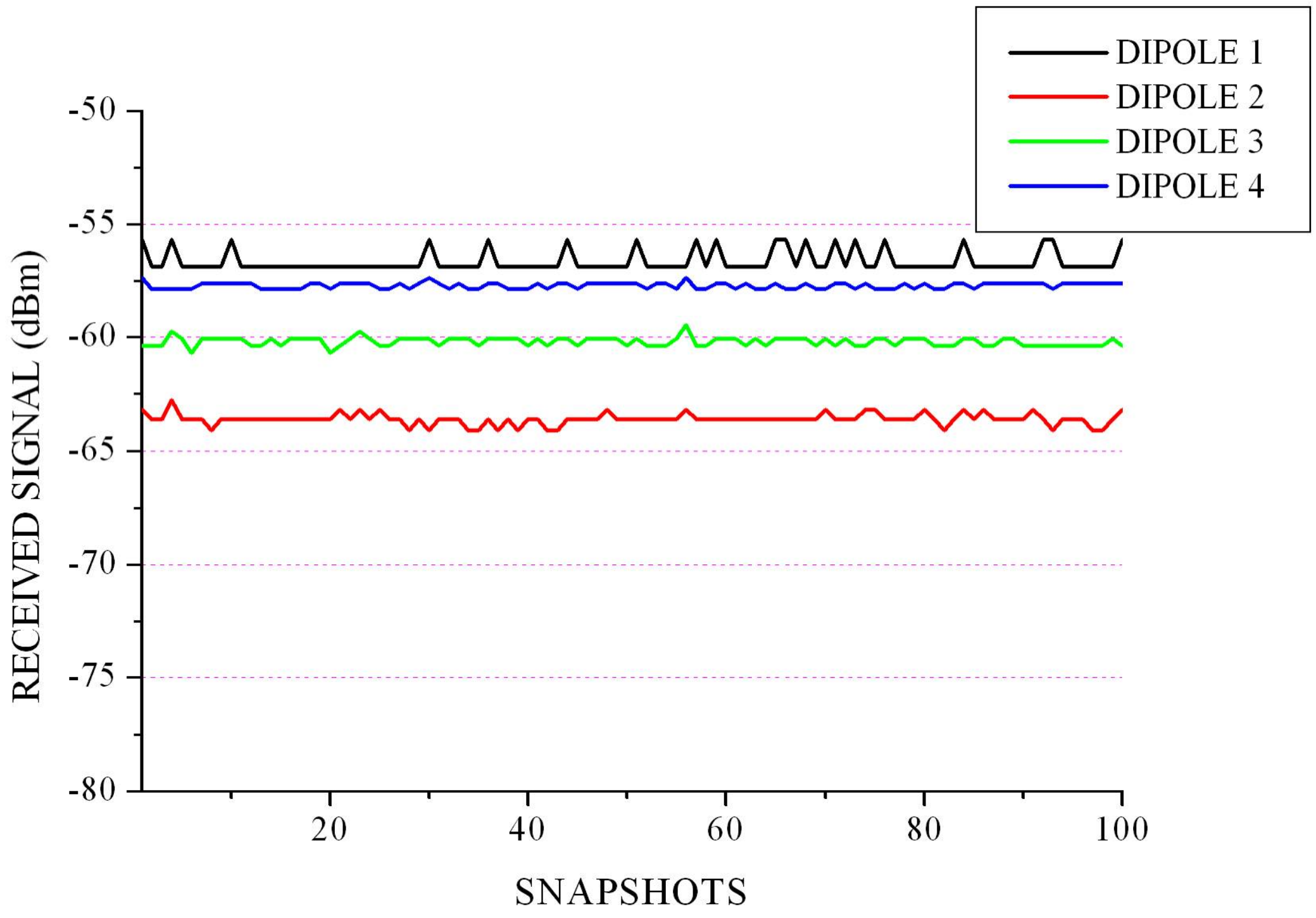

Figure 8. Received signal strength on ULA at the second time interval

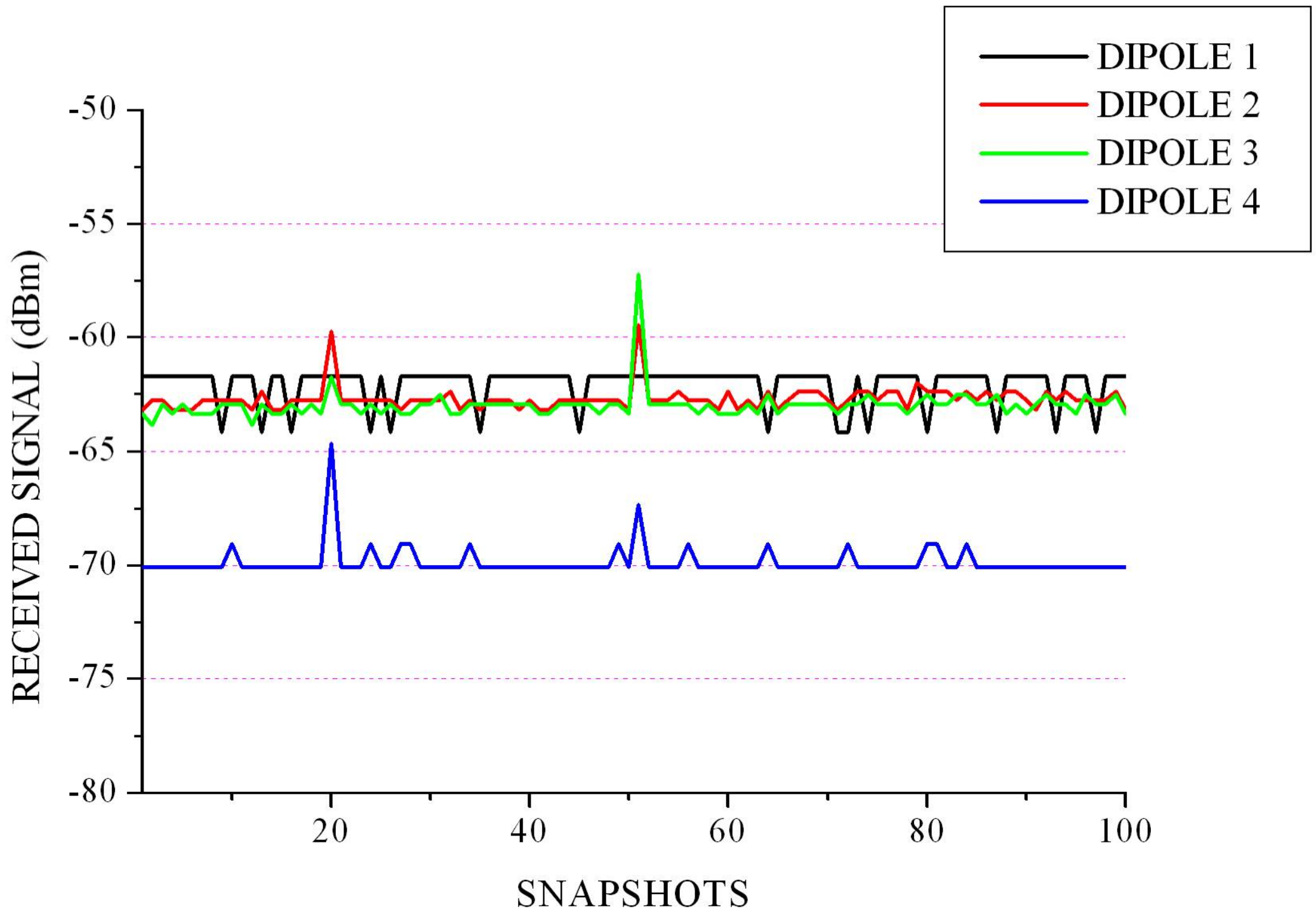

Figure 9. Received signal strength on Π-shape array at the first time interval

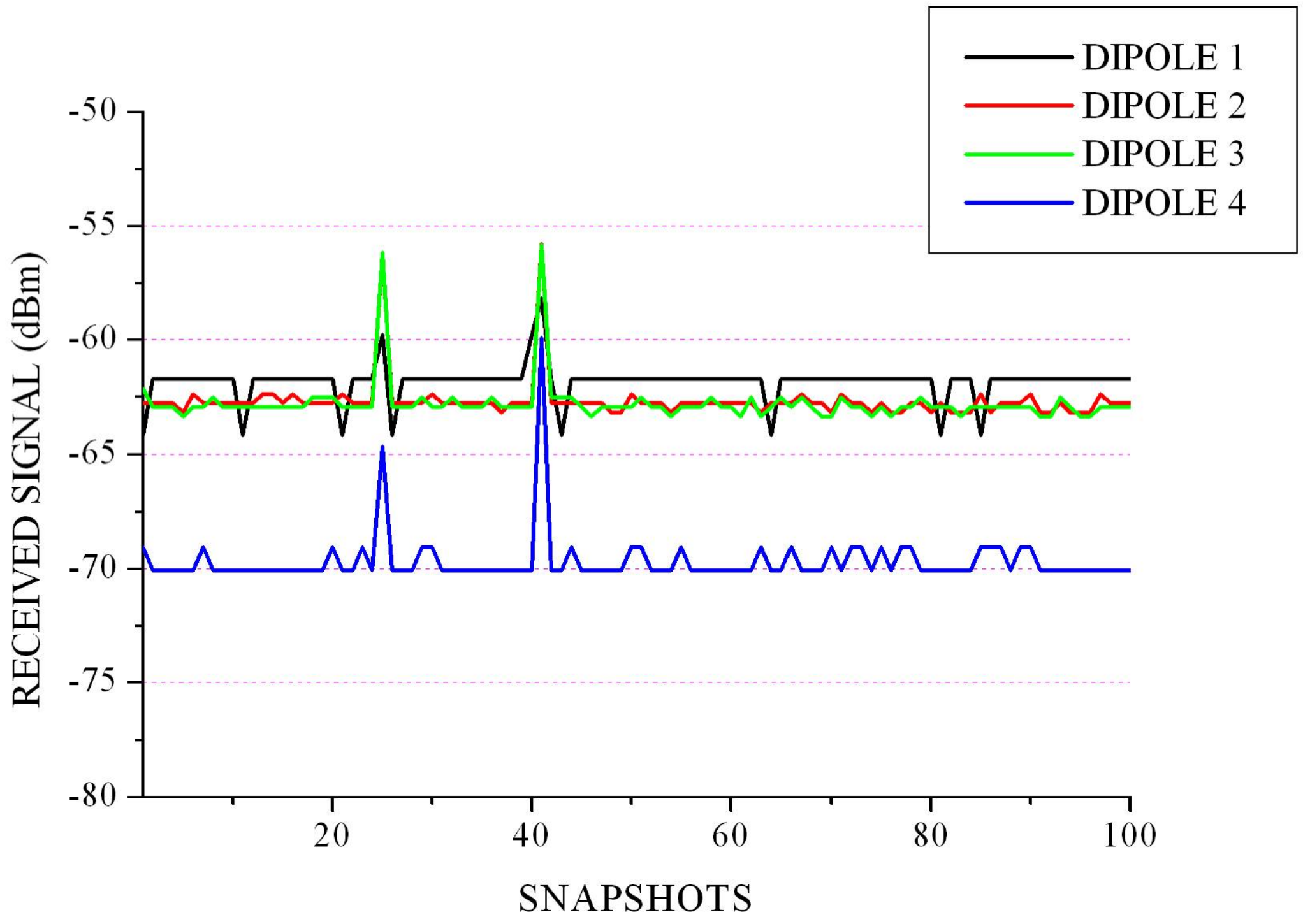

Figure 10. Received signal strength on Π-shape array at the second time interval

On both channel sounder application scenarios and at each antenna element, it is observed that the measured received signal strength was temporally varied. For the case of ULA configuration the power signal varies around 1-2 dB. For the Π-shape configuration there are quite wider range power signal variations that are on the order of 5-10 dB. The received power signal strength variations, at each ULA antenna element, are quite uncorrelated. This consideration is also highlighted in the case of Π-shape antenna array configuration. There are limited received signal strength declinations, between ULA and Π-shape configurations that are possibly provided by two distinct reasons. The first reason may be described by the fact that the four dipole elements at receiving ULA array configuration have the same polarization so that the axes of their arms are arranged in parallel. Also, these axes are in parallel with the transmitting dipole arms. Instead, the Π-shape antenna array first and fourth dipoles are arranged on side by side configuration and their polarizations are perpendicular to the dipole element transmitter antenna polarization. The second reason seems to be related to the fact that there is a quite strong secondary signal replica at the receiver side. The replica signal is mainly received by the first dipole at Π-shape configuration. Instead, this signal replica seems to be blocked by the geometry of Π-shape antenna array. For this reason, the obtained signal strength results indicate a quite low receiving signal level at the fourth dipole. The blocking of these signal components may provide quite wide received signal strength temporal variations on fourth dipole at Π-shape antenna array topology. For both ULA and Π-shape antenna array configurations the distances between the antenna elements are at around one wavelength at the 2.4GHz frequency band which ensures uncorrelated receiving power variations among elements. Using either receiving ULA or Π-shape antenna array, the static radio propagation environment and the light of sight path existence ensure quite limited received power variations. These variations are larger in case of using Π-shape receiving antenna array geometry. In general, the received signal strength declinations at ULA or Π-shape configuration partially depend on the receiving antenna array geometry. Also, the under test arrays introduce antenna polarization aspects that seem to have additional impact on the received signal strength variations.

### *3.2.2 Normalized channel gain coefficients*

For further investigation, the normalized channel gain coefficients $K_{i1}$, were calculated for both geometry configurations. Index i corresponds to one of the four sub-channels co-existing at 1x4 SIMO indoor radio propagation environment. The normalization process, on channel gain coefficients, was achieved using the first sub-channel gain coefficient values as reference. The first sub-channel was defined by a single element transmitting antenna and the first of the four element receiving antenna array. Figures 11 and 12 plot the normalized channel gain coefficients for the ULA configuration at the first and the second time interval, respectively. Figures 13 and 14 depict the normalized channel gain coefficients in case of using the Π- configuration.

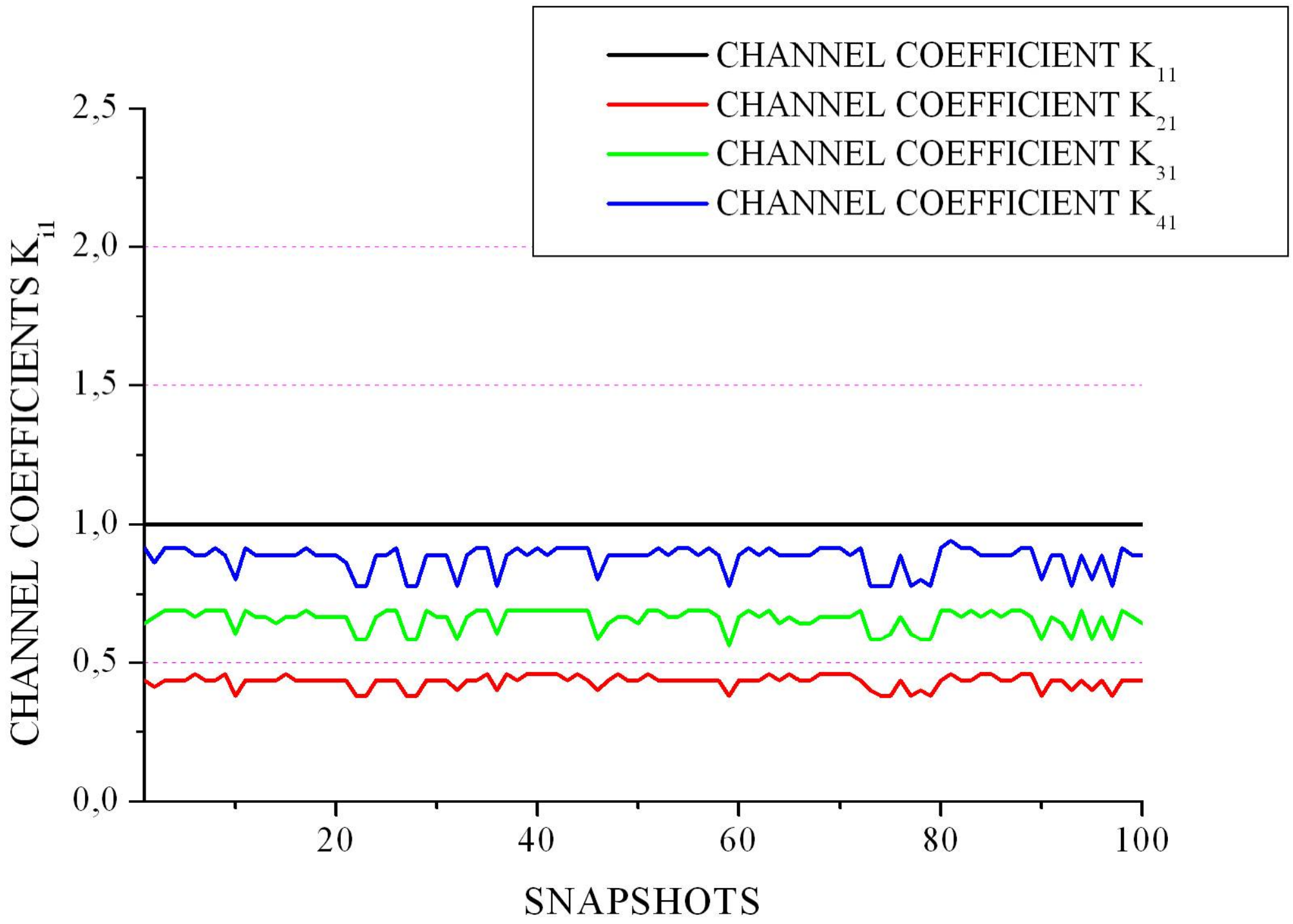

Figure 11. ULA-Normalized channel gain coefficients on 1x4 SIMO system at first time interval

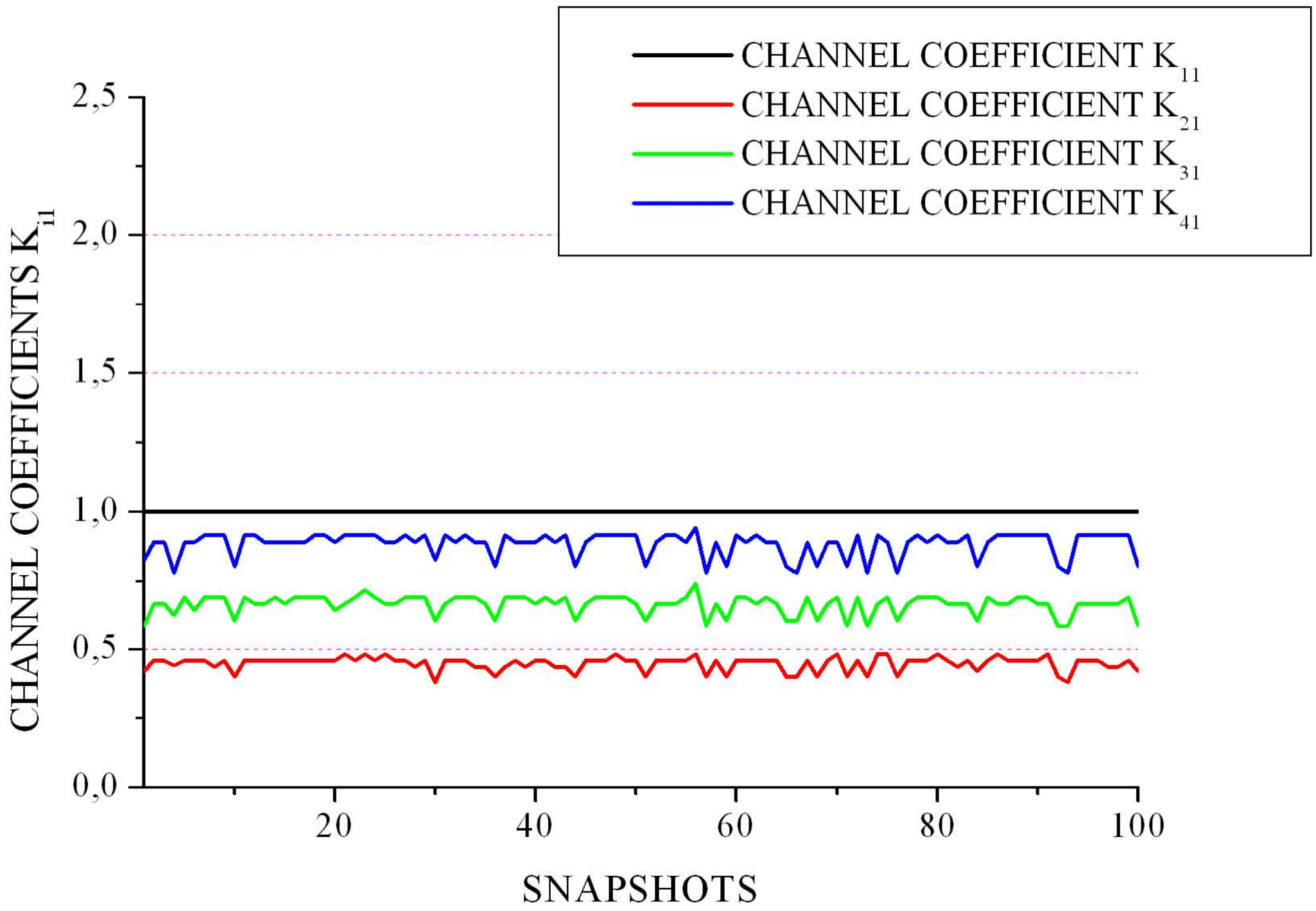

Figure 12. ULA-Normalized channel gain coefficients on 1x4 SIMO system at second time interval

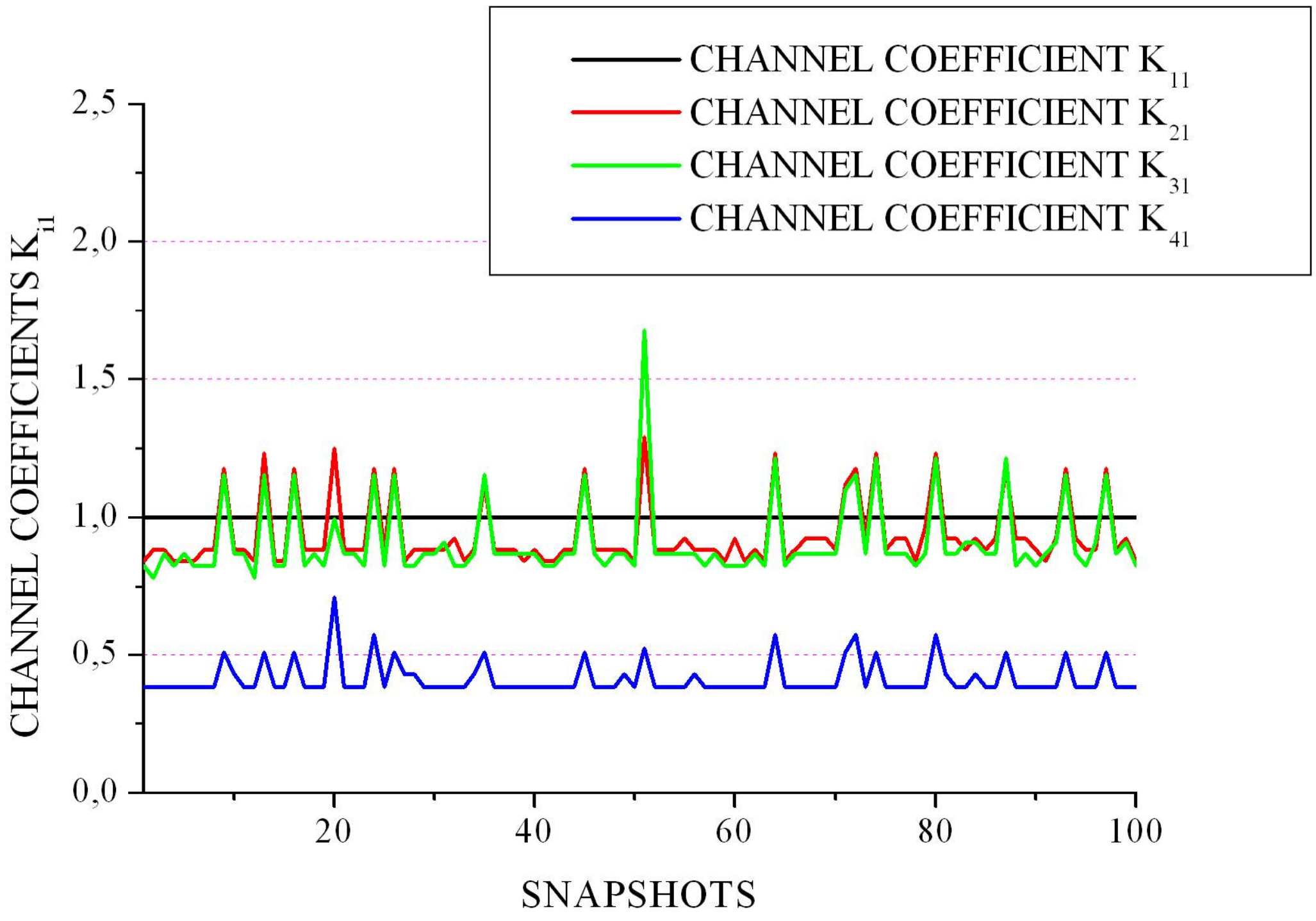

Figure 13. Π-shape array-Normalized channel gain coefficients on 1x4 SIMO system at first time interval

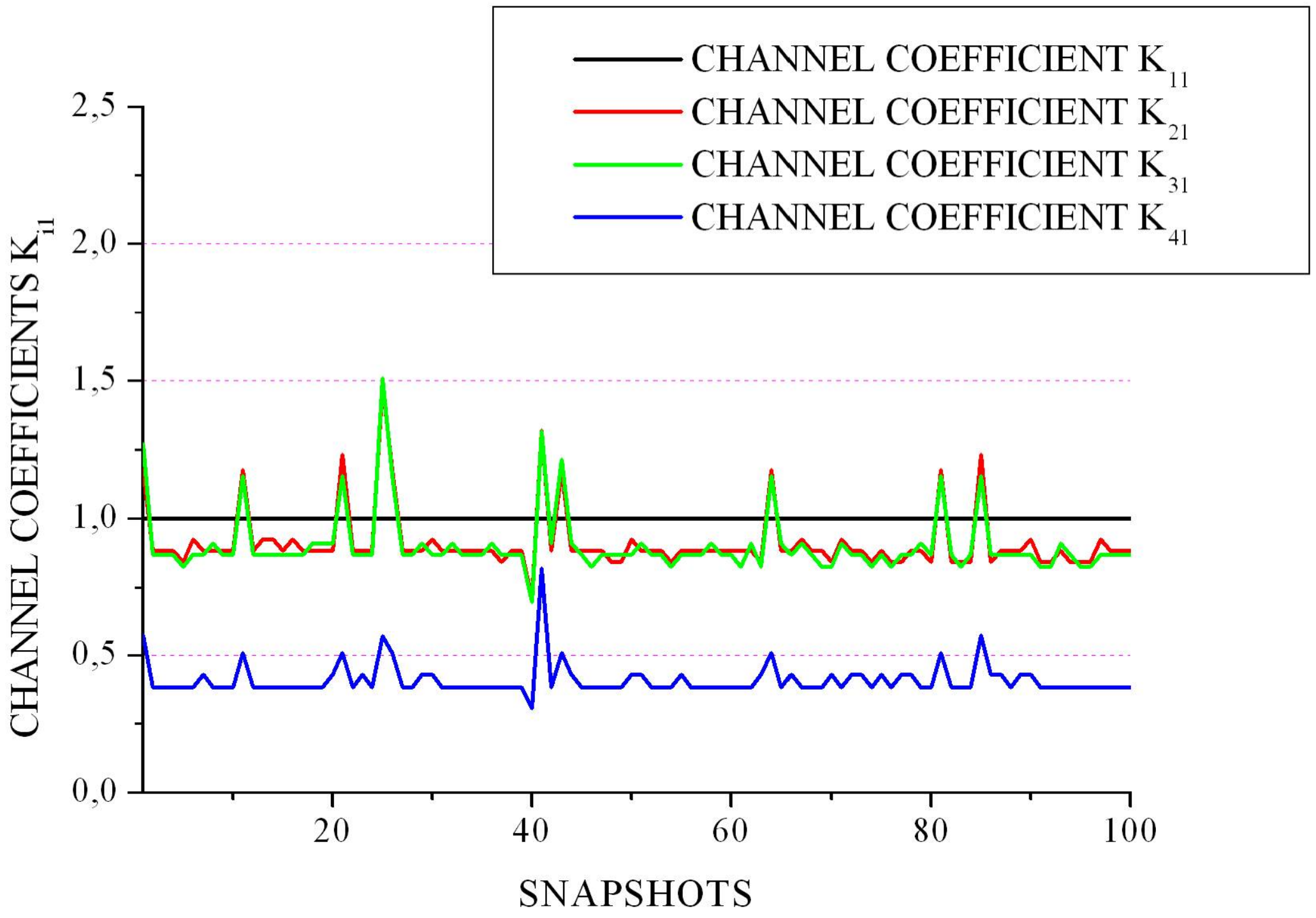

Figure 14. Π-shape array-Normalized channel gain coefficients on 1x4 SIMO system at second time interval

On ULA receiver testbed channel sounder measurements, the $K_{41}$ channel gain coefficient values are higher than the values of channel gain coefficients $K_{21}$ and $K_{31}$. This indicates that the sub-channel from single element transmitting dipole to the fourth dipole at the receiver antenna array exploits minor declination from the $K_{11}$ channel gain coefficient values. Instead, the sub-channel gain coefficient that provides major declination from $K_{11}$ values corresponds to the propagation path from single element transmitter dipole to the second dipole of receiver antenna array ($K_{21}$). The $K_{31}$ sub-channel coefficients on the 1x4 SIMO wireless channel ranges between the $K_{41}$ and $K_{21}$ thresholds.

In the case of using Π-shape antenna array on the receiver side, the $K_{21}$ and $K_{31}$ channel gain coefficients values exhibit quite equal values. Moreover, these channel gain coefficients present similar temporal variations. This attribute leads to the observation that these sub-channels are highly correlated. Moreover, the $K_{21}$ and $K_{31}$ channel gain coefficients values show quite limited declination from the $K_{11}$ values. The sub-channel signal path from single element transmitting dipole to the fourth dipole of receiving antenna array exhibits major declination from the $K_{11}$ values, providing the lowest $K_{i1}$ channel gain coefficient values.

The observations mentioned above seem to be provided by the fact that the Π-shape receiving antenna array geometry introduces polarization declinations amongst the corresponding dipole-elements. Hence, the differences on geometry aspects between the two receiving antenna configurations provide limited declinations on the range of channel gain coefficients. This attribute is shown on received signal power curves, too.

*3.2.3 Capacity evaluation*

For channel capacity evaluation the measured channel matrix gain coefficient values and the mathematical expression of equation (1) were used. Figure 15 shows the estimated channel capacity for ULA and Π-shape receiving antenna array at first time interval. Channel capacity estimation was also provided using the channel sounder results at second time interval (Figure 16).

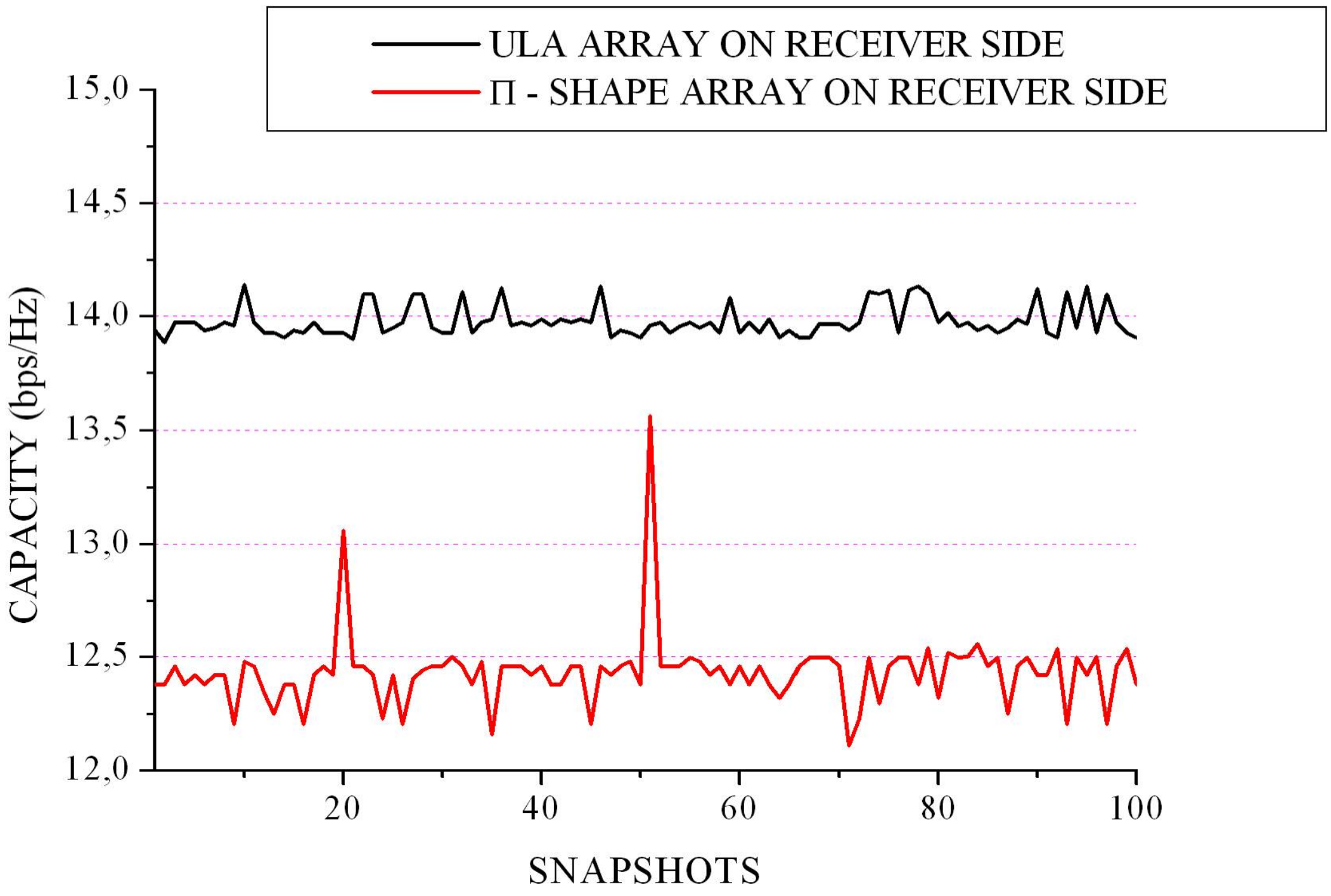

Figure 15. Estimated channel capacity on 1x4 SIMO system for ULA and Π-shape configuration at receiver side (first time interval)

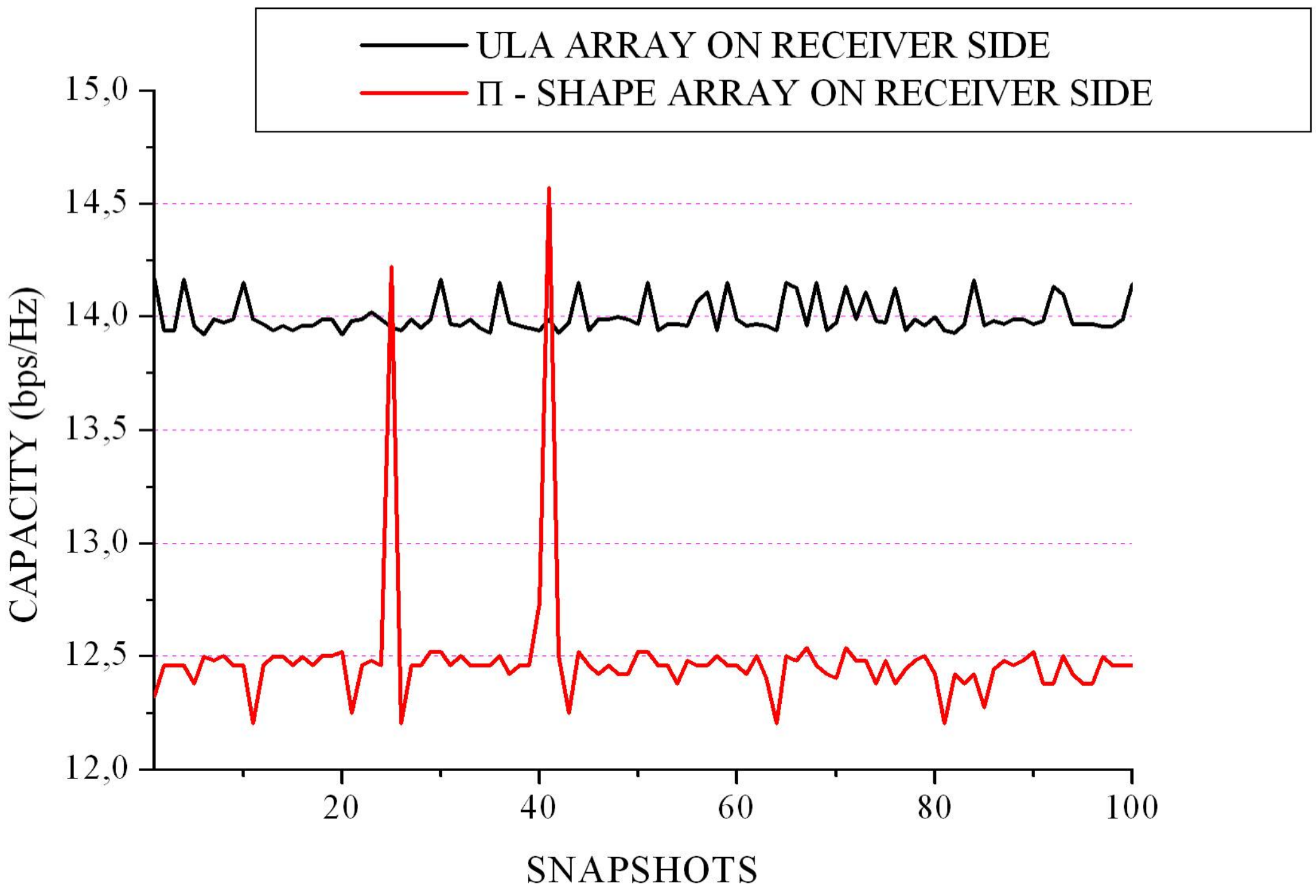

Figure 16. Estimated channel capacity on 1x4 SIMO system for ULA and Π-shape configuration at receiver side (second time interval).

Estimated channel capacity curves are indicating temporal variations on estimated channel capacity which are larger for the Π-shape receiving array than the ULA. For both receiving antenna array configurations, the temporal variations are mainly due to the slow fading radio propagation environment. Moreover, these channel capacity variations seem to be produced from the dipole's arrangement at two receiving antenna array topologies. These are also extended in case of Π-shape receiving antenna array scenario due to the temporal variations of received signal strength at first dipole that are provided by a quite strong signal replica existence. The channel capacity for ULA at receiver side approximates 14 bps/Hz. In the case of Π-shape antenna array at receivers end, the channel capacity value is close to 12.5 bps/Hz. It seems that 1x4 SIMO system with Π-shape receiving antenna array is less efficient in terms of channel capacity than the corresponding system with ULA receiving antenna array. The channel capacity values on both cases provide a limit on the bit rate that could be achieved on these wireless systems at the same radio propagation environment, remaining the bit error rate value at quite tolerable range. For indoor radio propagation environment, using the ULA at receiver side, the upper limit on the bit rate approximates to 14 bit/sec/Hz for efficient and reliable wireless communication applications. Replacing the ULA with the Π-shape antenna configuration, this upper limit is decreased at about 1.5 bit/sec/Hz.

For the investigation on channel capacity enhancement due to SIMO architecture, the normalization method that is based on equation (2) was used. From the measured results on channel matrix gain coefficients, the normalized channel capacity $C_n$ at ULA and Π-shape receiver antenna arrays at first and second time intervals was calculated. The results are shown in Figures 17 and 18.

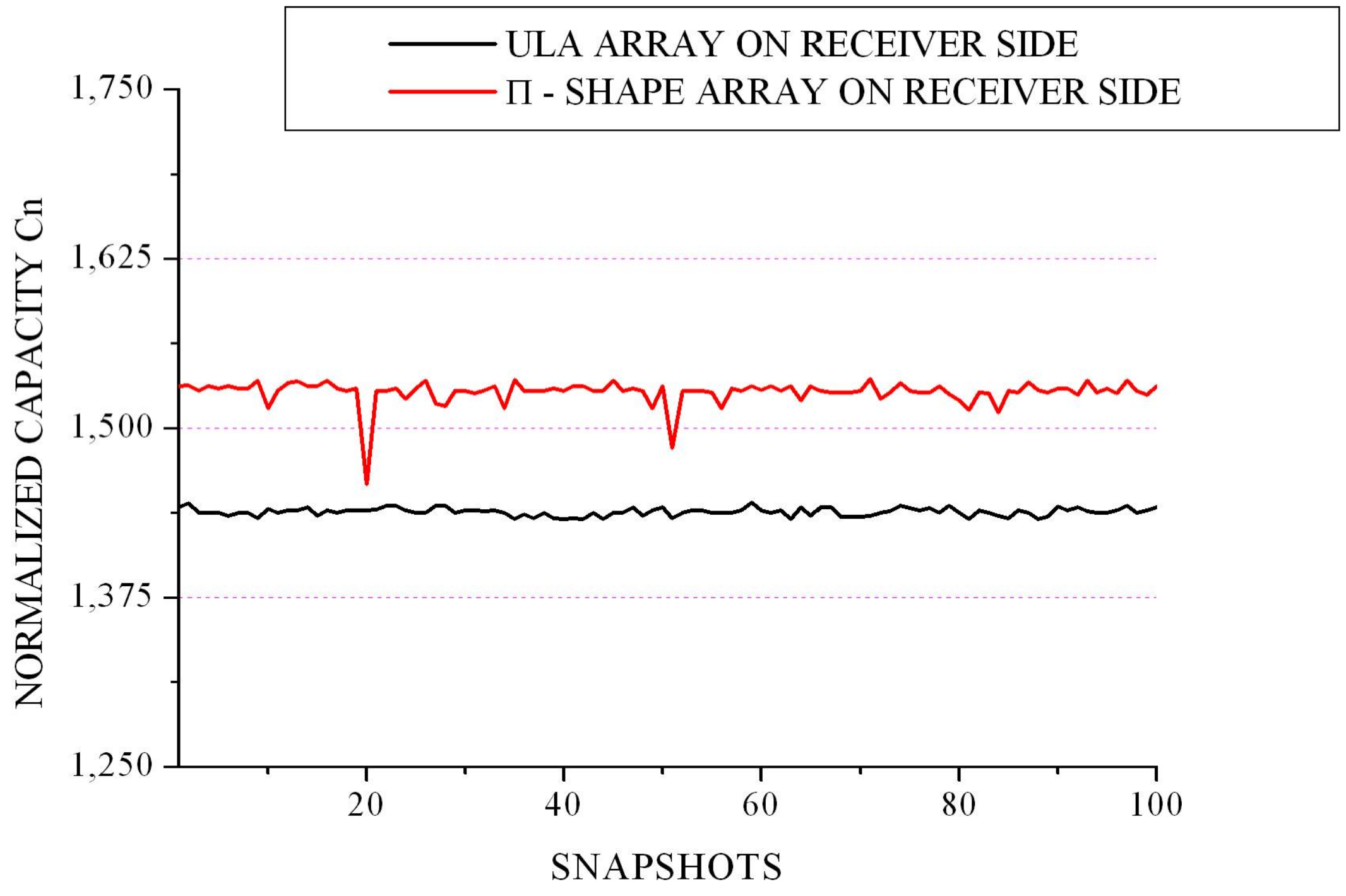

Figure 17. Normalized channel capacity on 1x4 SIMO system for ULA and Π-shape configuration at receiver side (first time interval)

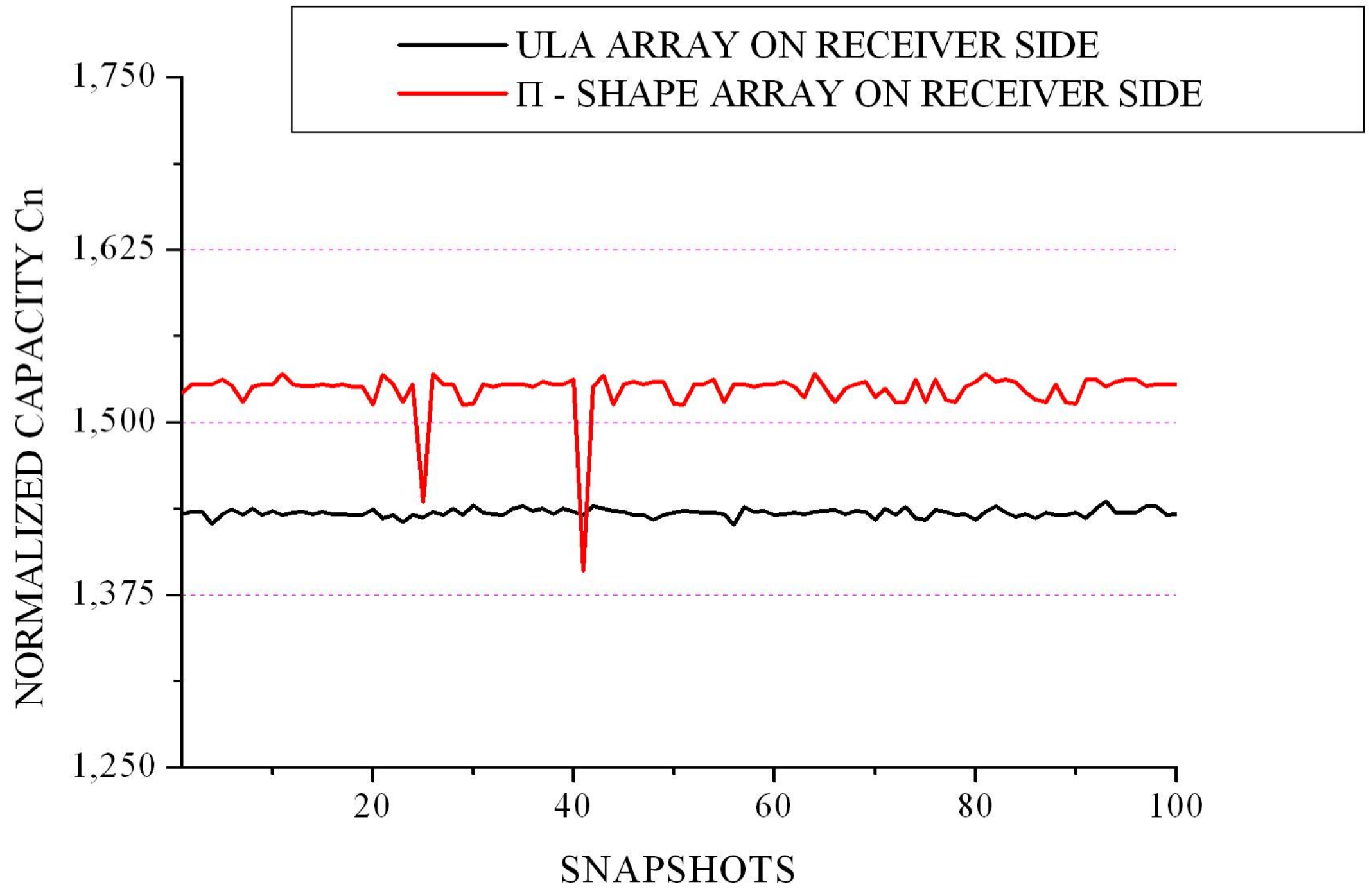

Figure 18. Normalized channel capacity on 1x4 SIMO system for ULA and Π-shape configuration at receiver side (second time interval)

From these curves, the normalized capacity parameter $C_n$ indicates limited time variations due to slow fading static radio propagation environment. These temporal variations are wider in the case of using Π-shape receiving antenna array. As mentioned above, the Π-shape antenna array topology in conjunction with the radio propagation environment allows channel capacity temporal variations. Both the ULA and Π-shape receiving antenna array configurations provide channel capacity enhancement that approximate the limit of 1.5. For ULA on the receiver end, the 1x4 SIMO channel capacity is 1.44 times higher than the averaged channel capacity over the four SISO sub-channels. This capacity enhancement is more efficient in case of Π-shape configuration than on the averaged channel capacity over the four SISO sub-channels. In fact, the value of the normalized channel capacity in case of using the Π-shape receiving antenna array approximates to 1.52. In any case of receiving antenna array geometries, the obtained results provide the evaluation of channel capacity enhancement, due to SIMO architecture utilization

*3.2.4 Comparative Study*

Since all these type of metrics are strongly depended on a variety of parameters, such as: measurement environment, testbed configuration, type and numbers of the antennas, narrowband/wideband assumption, signal to noise ratio, distance between the transmitter and the receiver, it is extremely difficult to have fair comparison with previous works. On the other hand in this subsection, similar multiple elements channel sounder test-beds are presented, and comparative studies are given whenever possible.

A 1X4 SIMO platform using UWB prototype antenna elements linearly arranged at inter-element distance separation 85mm over ground planes was implemented by the Atomic Energy Commission Laboratory (Keignart, Abou-Rjeily, Delaveaud, & Daniele 2006). Concerning antenna array measurements in an anechoic chamber at 3.8GHz center frequency experimental results are shown received signal in 2 middle elements around -1.2dB compared to the side antennas. In ETA Lab indoor laboratory environment measurements at 2.4GHz are shown differences around 1, 2 and 7dB between either linearly or Π-shape arranged elements. Differences that explained by diffraction array effect (Hao Yuan, Hirasawa and Yimin Zhang, 1998) and indoor measurement environment.

Another multi-element channel sounder developed and characterized at Brigham Young University (Wallace & Jensen 2005). Antenna arrays at transmit and receive were 8-element uniform circular arrays, consisting of omnidirectional monopole elements with $\lambda/2$ inter-element spacing. The transmitter was stationary in a hallway, while the receiver was placed in 8 different rooms. The channel capacity for these eight rooms varies from 2bps/Hz to values larger than 20bps/Hz while the SNR was 20dB at the center frequency of 2.55GHz. In ETA Lab environment, the transmitter and the receiver were placed inside the same laboratory room at a distance 4.5m while the SNR was 33dB. The channel capacities were around 14 and 12.5bps/Hz for linear and Π-shape array respectively.

A research team at NTT Network Innovation Laboratories has evaluated the frequency efficiency by obtaining the bit error rate of a 16×16 Multiuser MIMO Testbed operating at 4.85GHz in an actual indoor environment (Nishimori et al., 2010). Linear array configuration was used for both transmitter and receiver side at a minimum distance around 5m. The frequency utilization of 43.5bps/Hz and 50 bps/Hz (1Gbps) was achieved when the SNRs were 31 and 36dB, respectively.

Extensive channel measurements were performed at the center carrier frequency of 5.25 GHz with 100 MHz bandwidth at Beijing University of Posts and Telecommunications (Nan, Zhang, Zhang & Lei, 2011). The capacity for 4x4 polarized antennas with λ/2 spacing between the antennas and 15m distance between the transmitter and the receiver was 26bps/Hz under LoS propagation.

From the previous paragraphs, the impact of a number of antenna elements and the distance between transmitter and the receiver to channel capacity is mainly highlighted.

## 4. Conclusion

An adaptable, cost-effective testbed measurement platform for SIMO channel sounder applications was designed and implemented. Two districts antenna array configurations were compared on the SIMO measurement set-up for channel capacity investigation and channel characterization on an indoor radio propagation environment. Complex channel gain coefficients and channel capacity estimation were accomplished. The calculated and estimated results firstly indicate the temporal signal strength variation at each receiver antenna element. Besides, the channel gain coefficients present quite limited correlation. These coefficients also exhibit limited temporal variations due to the slow fading static radio propagation environment. For both receiving antenna array configurations, the 1x4 SIMO technique offers capacity enhancement on the order of 1.5 regarding the averaged capacity of the corresponding four SISO sub-channels. Using the Π-shape receiving antenna array, 1x4 SIMO channel capacity has quite lower value than in the case of using ULA configuration on the receiver end. Instead, using the Π-shape antenna array, the normalized channel capacity values are higher than in the case of using the ULA configuration at receiver end. All the mentioned remarks are obviously very critical for indoor wireless communication systems on radio propagation environments.